\documentclass[twocolumn,showpacs,showkeys,superscriptaddress,floatfix]{revtex4}

\usepackage{graphicx}

\usepackage{epsfig}
\usepackage{amssymb}
\usepackage{amsmath}
\usepackage{amsfonts}
\usepackage{epstopdf}

\makeatother

\begin{document}

\title{Status of the lepton $g-2$ and effects of hadronic corrections}
\author{Alexander E. Dorokhov}\email{dorokhov@theor.jinr.ru}
\affiliation{Joint Institute for Nuclear Research, Bogoliubov
  Laboratory of Theoretical Physics, 114980, Dubna, Russia;\\
N.N.Bogoliubov Institute of Theoretical Problems of Microworld,
M.V.Lomonosov Moscow State University, Moscow 119991, Russia  }
\author{Andrei E. Radzhabov}\email{aradzh@icc.ru}
 \affiliation{Institute for System Dynamics and Control Theory SB RAS, 664033 Irkutsk,
Russia  }\author{Alexei S. Zhevlakov}\email{zhevlakov1@gmail.com}
 \affiliation{Institute for System Dynamics and Control Theory SB RAS, 664033 Irkutsk,
Russia  }
\date{\today}

\begin{abstract}
The electron and muon anomalous magnetic moments (AMM) are measured in experiments and studied in the Standard Model (SM) with the highest precision accessible in particle physics. The comparison of the measured quantity with the SM prediction for the electron AMM provides the best determination of the fine structure constant. The muon AMM is more sensitive to the appearance of New Physics effects and, at present, there appears to be a three- to four-standard deviation between the SM and experiment. The lepton AMMs are pure relativistic quantum correction effects
and therefore test the foundations of relativistic quantum field theory in general, and of quantum electrodynamics (QED) and SM in particular, with highest sensitivity. Special attention is paid to the studies of the hadronic contributions to the muon AMM which constitute the main source of theoretical uncertainties of the SM.
\end{abstract}


\centerline{\it Dedicated to the memory of Professor Eduard Alekseevich Kuraev}

\maketitle
\section{Motivation}

\begin{figure}[h]
\begin{center}
\includegraphics[height=6cm]{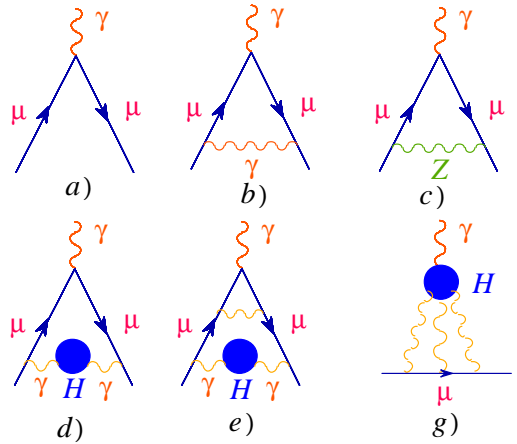}%
\end{center}
\caption{Representative diagrams for the SM contributions to $a_\mu$.
Here, $H$ is for the hadronic block.}
\label{SM}
\end{figure}

The anomalous magnetic moment (AMM) of charged leptons ($l=e,\mu ,\tau $) is
defined by%
\begin{equation}
a_{l}=\frac{g_{l}-2}{2},  \label{a}
\end{equation}%
with the gyromagnetic ratio $g_{l}$ of the lepton magnetic moment to its
spin, in Bohr magneton units. For a free pointlike fermion one has $g=2$
in accordance with the Dirac equation (Fig.~\ref{SM}a). However, deviations
appear when taking into account the interactions leading to fermion substructure and
thus to nonzero $a_{l}$ (Fig.~\ref{SM}b-g).

During the first years of the lepton AMM studies the fundamental task was to
test the foundations of quantum field theory in general and quantum
electrodynamics (QED) in particular. At present, the measurements of the
lepton AMM are one of the major low-energy tests of the standard model (SM)
and play an important role in the search for new interactions beyond the SM.

The nonzero lepton AMMs are induced by radiative corrections due to the coupling
of the lepton spin to virtual fields, which in the SM are induced by QED, weak and
strong (hadronic) interactions\footnote{%
For comprehensive reviews see \cite{Jegerlehner:2009ry,Miller:2012opa}.}
(Fig. ~\ref{SM})%
\begin{equation}
a^{\mathrm{SM}}=a^{\mathrm{QED}}+a^{\mathrm{weak}}+a^{\mathrm{hadr}}.
\label{aSM}
\end{equation}%
\qquad\ The electron and muon AMMs are among the most accurately measured
quantities in elementary particle physics. Today, the electron AMM serves as
the best quantity to determine the fine structure constant with the highest
accuracy. At the same time, for $a_\mu$, there is a deviation at the level of 3-4 $\sigma $ of
the SM prediction from the measured value. Even if this does
not give a clear indication for the existence of New Physics, it allows us to provide
stringent constraints on the parameters of hypothetical models.

The outline of this paper is as follows. In Section 2, we briefly report on
the status of the electron AMM. In Section 3, the latest experimental and
theoretical results on $a_\mu$ are presented. In Section 4, we review
the most problematic theoretical input coming from the contribution due to
the hadronic light-by-light (HLbL) scattering mechanism. Section 5 is
devoted to some technical details of the calculations of the HLbL within the
nonlocal chiral quark model (N$\chi $QM). Conclusions are presented in
the last Section.

\section{Electron AMM and fine structure constant}

In 2008, the unique measurement by the Harvard group of Prof. G. Gabrielse,
using a one-electron quantum cyclotron, obtained the electron AMM with
unprecedented accuracy \cite{Hanneke:2008tm}
\begin{equation}
a_{e}^{\mathrm{Harvard}}=1~159~652~180.73~(0.28)\times 10^{-12}\quad \lbrack
0.24~\mathrm{ppb}].  \label{aeHarv}
\end{equation}%
This result leads to the determination of the fine structure constant $\alpha $
with the extraordinary precision \cite{Aoyama:2012wj,KINOSHITA:2014uza}%
\begin{equation}
\alpha ^{-1}=137.035~999~1727~(341)\quad \lbrack 0.25~\mathrm{ppb}].
\label{alpha_QED}
\end{equation}%
The latter became possible after the complete QED contribution to the
electron AMM up to tenth order in the coupling constant were achieved
numerically by the Prof. T. Kinoshita group \cite{Aoyama:2012wj} (for recent
review see \cite{KINOSHITA:2014uza}). Note, that the uncertainty in (%
\ref{alpha_QED}) is dominated by the uncertainty in the measurement of $a_{e}^{\mathrm{Harvard}}$.

The value in (\ref{alpha_QED}) has the highest precision of any value of $%
\alpha $ currently available \cite{Mohr:2012tt}. Thus, a new measurement
\cite{Bouchendira:2010es} of the ratio $h/m_{\mathrm{Rb}}$ between the
Planck constant and the mass of $^{87}\mathrm{Rb}$ atom studied by atom
recoil leads to a value of the fine structure constant \cite{Mohr:2012tt} $%
\alpha ^{-1}\left( \mathrm{Rb}\right) =137.035~999~049~(90)\quad \lbrack
0.66~\mathrm{ppb}].$ By using this $\alpha $ one gets for the electron AMM%
\begin{equation}
a_{e}^{\mathrm{SM}}\left( \mathrm{Rb}\right) =1~159~652~181.78~(0.77)\times
10^{-12}\quad \lbrack 0.66~\mathrm{ppb}],  \label{aeSM}
\end{equation}%
that is in agreement with the measurement (\ref{aeHarv}).

\section{Muon AMM: experiment vs theory}

In 2006,  the results on the $a_\mu$ measurements by
the E821 collaboration at Brookhaven National Laboratory \cite%
{Bennett:2006fi} were published. The combined result, based on nearly equal samples of
positive and negative muons, is\footnote{%
Lateron this value was corrected \cite{Mohr:2012tt} for a small shift in
the ratio of the magnetic moments of the muon and the proton as $a_{\mu }^{%
\mathrm{BNL,corr}}=116~592~09.1~(6.3)\times 10^{-10}.$}%
\begin{equation}
a_{\mu }^{\mathrm{BNL}}=116~592~08.0~(6.3)\times 10^{-10}\quad \lbrack 0.54~%
\mathrm{ppm}].  \label{amuBNL}
\end{equation}%
This exiting result is still limited by the statistical errors and proposals
to measure $a_\mu$ with a fourthfold improvement in accuracy have been
proposed at Fermilab (USA) \cite{Venanzoni:2012sq} and J-PARC (Japan) \cite%
{Saito:2012zz}. A future experiments plan to reduce the present experimental
error to a precision of $0.14$ ppm.

In the SM the dominant contribution to the lepton AMM comes from QED (Fig.~\ref%
{SM}b). The complete tenth-order QED contribution to $a_\mu$ was
reported in \cite{Aoyama:2012wk}%
\begin{equation}
a_{\mu }^{\mathrm{QED}}=11~658~471.8951~(0.0080)\times 10^{-10}.
\label{aQED}
\end{equation}%
The accuracy of these calculations is enough for any planed experiments in
new future.

In general, the weak contributions (Fig.~\ref{SM}c) are small due to
suppressing factor $\alpha /\pi \cdot m_{\mu }^{2}/M_{w}^{2}\sim 10^{-9}$,
where $M_{w}$ is a typical mass of heavy $W^{\pm },Z$ and $H$ bosons.
The one- and two-loop evaluations indicate that they are known with a
sufficiently high accuracy \cite{Czarnecki:2002nt,Gnendiger:2013pva}%
\begin{equation}
a_{\mu }^{\mathrm{weak}}=15.36~\left( 0.10\right) \times 10^{-10},
\label{aWeak}
\end{equation}%
where the remaining theory error comes from the unknown three-loop contributions
and dominantly from light hadronic uncertainties in the second-order
electroweak diagrams with quark triangle loops. The most important feature
of these new estimates, that significantly increase the theoretical
precision, is to use the LHC result on the Higgs-boson mass measured by
ATLAS \cite{Aad:2012tfa,Aad:2013wqa} and CMS \cite%
{Chatrchyan:2012ufa,Chatrchyan:2012jja} Collaborations.

Strong (hadronic) interaction produces relatively small contributions to $a_\mu$, however they are known with an accuracy comparable to the
experimental uncertainty in (\ref{amuBNL}). In leading in $\alpha$ orders, these contributions can be
separated into three terms%
\begin{equation}
a_{\mu }^{\mathrm{hadr}}=a_{\mu }^{\mathrm{HVP}}+a_{\mu }^{\mathrm{ho}%
}+a_{\mu }^{\mathrm{HLbL}}.  \label{aStrong}
\end{equation}%
In (\ref{aStrong}), $a_{\mu }^{\mathrm{HVP}}$ is the leading in $\alpha $ contribution due to the hadron vacuum
polarization (HVP) effect in the internal photon propagator of the one-loop
diagram (Fig.~\ref{SM}d), $a_{\mu }^{\mathrm{ho}}$ is the next-to-leading
order contribution related to iteration of HVP (Fig.~\ref{SM}e). The last
term is not reduced to HVP iteration and it is due to the hadronic
light-by-light (HLbL) scattering mechanism (Fig.~\ref{SM}g).

The hadronic contributions in (\ref{aStrong}) are determined by effects
dominated by the long distance dynamics, the region where the methods of
perturbation theory of Quantum Chromodynamics (QCD) do not applicable and
one must use less reliable nonperturbative approaches. However, in case of
HVP, using analyticity and unitarity (the optical theorem) $a_{\mu }^{%
\mathrm{HVP}}$ can be expressed as the spectral representation integral \cite%
{BM61}
\begin{equation}
a_{\mu }^{\mathrm{HVP}}=\frac{\alpha }{\pi }\int_{4m_{\pi }^{2}}^{\infty }%
\frac{dt}{t}K(t)\rho _{\mathrm{V}}^{(\mathrm{H})}\left( t\right) ,
\label{Amm_rho}
\end{equation}%
which is a convolution of the hadronic spectral function
\begin{equation}
\rho _{V}^{\mathrm{(H)}}\left( t\right) =\frac{1}{\pi }\mathrm{Im} \Pi ^{\mathrm{(H)%
}}\left( t\right)  \label{rhoH}
\end{equation}%
with the known from QED kinematical factor%
\begin{equation}
K(t)=\int_{0}^{1}dx{\frac{x^{2}(1-x)}{x^{2}+(1-x)t/m_{\mu }^{2}}},
\label{Kfac}
\end{equation}%
where $m_{\mu }$ is the muon mass. The QED factor is sharply peaked at low
invariant masses $t$ and decreases monotonically with increasing $t$. Thus,
the integral defining $a_{\mu }^{\mathrm{HVP}}$ is sensitive
to the details of the spectral function $\rho _{V}^{\mathrm{(H)}}\left(
t\right) $ at low $t$. At present there are no direct theoretical tools that
allow one to calculate the spectral function at low $t$ with required accuracy.
Fortunately, $\rho _{V}^{\mathrm{(H)}}\left( t\right) $ is related to the
total $e^{+}e^{-}\rightarrow \gamma ^{\ast }\rightarrow $ hadrons
cross-section $\sigma (t)$ at center-of-mass energy squared $t$ by
\begin{equation}
\sigma ^{e^{+}e^{-}\rightarrow \mathrm{hadrons}}(t)=\frac{4\pi \alpha }{t}%
\rho _{\mathrm{V}}^{(\mathrm{H})}\left( t\right) ,  \label{Sigma_rho}
\end{equation}%
and this fact is used to get quite accurate estimate of $a_{\mu }^{\mathrm{%
HVP}}$. The most precise recent phenomenological evaluations of $a_{\mu }^{%
\mathrm{HVP}}$, using recent $e^{+}e^{-}\rightarrow \mathrm{hadrons}$ data,
provide the results%
\begin{equation}
a_{\mu }^{\mathrm{HVP,}e^{+}e^{-}}=\left\{
\begin{array}{l}
692.3~\left( 4.2\right) \times 10^{-10},\quad \text{\cite{Davier:2010nc}} \\
694.91~\left( 4.27\right) \times 10^{-10}.\quad \text{\cite{Hagiwara:2011af}}%
\end{array}%
\right.  \label{aHVPd}
\end{equation}%
In addition, the data on inclusive decays of the $\tau $-lepton into hadrons are
used to replace the $e^{+}e^{-}$ data in certain energy regions. This is
possible, since the vector current conservation law relates the $I=1$ part
of the electromagnetic spectral function to the charged current vector
spectral function measured in $\tau \rightarrow \nu $ +non-strange hadrons (see, i.e. \cite%
{Jegerlehner:2011ti}).
All these allow one to reach a substantial improvement
in the accuracy of the contribution from the HVP during the last decade.

Similarly, the dispersion relation approach and the same phenomenological input
lead to the estimate of the next-to-leading hadronic contribution (Fig.~\ref%
{SM}e) \cite{Hagiwara:2011af}%
\begin{equation}
a_{\mu }^{\mathrm{ho}}=-9.84~\left( 0.08\right) \times 10^{-10}.  \label{aHO}
\end{equation}%
Thus, the HVP and next-to-leading order contribution related to HVP are known with an
accuracy better than 1 per cent.

In near future, it is expected that new and precise measurements from CMD3
and SND at VEPP-2000 in Novosibirsk, BES III in Beijing and KLOE-2 at DAFNE
in Frascati allow one to significantly increase the accuracy of predictions for $%
a_{\mu }^{\mathrm{HVP}}$ and $a_{\mu }^{\mathrm{ho}}$ and resolve some
inconsistency problems between different set of data.

Combining all SM contributions one obtains
\begin{eqnarray}
&&a_{\mu }^{\mathrm{SM}}=116~591~80.2~(0.1)_{\mathrm{EW}}(0.08)_{\mathrm{ho}%
}(4.2)_{\mathrm{HVP}}(2.6)_{\mathrm{HLbL}}\notag\\
&&\times 10^{-10},  \label{aSMres}
\end{eqnarray}%
where we take the leading order evaluations given in (\ref{aHVPd}a) and
the guessed value for the hadronic light-by-light contribution from \cite%
{Prades:2009tw}%
\begin{equation}
a_{\mu }^{\mathrm{HLbL}}(\mathrm{Guess})=10.5~\left( 2.6\right) \times
10^{-10}.  \label{aLbLGuess}
\end{equation}%
The latter contribution will be discussed below with detail. The resulting
difference between the experimental result (\ref{amuBNL}) and the full SM
prediction is%
\begin{equation}
a_{\mu }^{\mathrm{BNL}}-a_{\mu }^{\mathrm{SM}}=28.7~\left( 8.0\right) \times
10^{-10},  \label{aDif}
\end{equation}%
which signals an $3.6~\sigma $ discrepancy between theory and experiment.
The SM theoretical error is dominated by the hadronic contributions. In that
respect, theoretical predictions of HVP and HLbL contributions to $a_\mu$ should be of the same level or better than the precision of planed
experiments.

\section{Status of the hadronic light-by-light scattering contribution to
$a_\mu$}

The next-to-leading order corrections are suppressed by the absolute value
by an extra degree of $\alpha $. However, one kind
of these contributions, corresponding to the HLbL (Fig.~\ref{SM}g), is of the amount
ranging from 0.5 to 1.5 ppm and known with accuracy of order 50\%. It
gives an error comparable in magnitude with the uncertainty induced by HVP (%
\ref{aHVPd}). The problem is that the HLbL scattering contribution can not be calculated from first principles
or (unlike to HVP) directly extracted from phenomenological considerations. Instead, it has to be evaluated using
various QCD inspired hadronic models that correctly reproduce low- and
high- energy properties of the strong interaction. Nevertheless, as will be
discussed below, it is important for the model calculations that
phenomenological information and well established theoretical principles
should significantly reduce the number of model assumptions and the
allowable space of model parameters.

In general, the HLbL scattering amplitude is a complicated object for
calculations. It is the sum of different diagrams including the dynamical
quark loop, the meson exchanges, the meson loops and the iterations of these
processes. Fortunately, already in the first papers devoted to the
calculation of the HLbL contributions \cite%
{deRafael:1993za,Hayakawa:1995ps,Bijnens:1995cc}, it has been recognized
that these numerous terms show a hierarchy. This is related to the existence of
two small parameters: the inverse number of colors $1/N_{c}$ and the ratio
of the characteristic internal momentum to the chiral symmetry parameter $%
m_{\mu }/(4\pi f_{\pi })\sim 0.1$. The former suppresses the multiloop
contributions, so that the leading contribution is due to the quark loop
diagram and the two-loop diagrams with mesons in the intermediate state.
In latter case, the contribution of the diagram with intermediate pion is enhanced by
small pion mass in the meson propagator.

Different approaches to the calculation of the
contributions from the HLbL scattering process to $a_\mu$ were used.
These approaches can be separated in several groups. The first one consists
of various extended versions of the vector meson dominance model (VMD)
supplemented by ideas of the chiral effective theory, such as the hidden
local symmetry model (HLS) \cite{Hayakawa:1995ps}, the lowest meson
dominance (LMD) \cite{Knecht:2001qf,Melnikov:2003xd,Nyffeler:2009tw}, the
resonance chiral theory (R$\chi $T) \cite{Kampf:2011ty,Roig:2014uja}. The
second group is based on the consideration of effective models of QCD that use
the dynamical quarks as effective degrees of freedom. The latter include
different versions of the (extended) Nambu--Jona-Lasinio model (E)NJL \cite%
{Bijnens:1995cc,Bartos:2001pg}, the Constituent Quark Models with local
interaction (CQM) \cite%
{Pivovarov:2001mw,Erler:2006vu,Boughezal:2011vw,Greynat:2012ww}, the models
based on nonperturbative quark-gluon dynamics, like the non-local chiral
quark model (N$\chi $QM) \cite{Dorokhov:2008pw,Dorokhov:2012qa}, the
Dyson-Schwinger model \cite{Fischer:2010iz} (DS), or the holographic models
(HM) \cite{Hong:2009zw,Cappiello:2010uy}. The lattice calculations of HLbL
are still at an exploratory stage \cite{Blum2013}.

The results of the model calculations are given in Tables \ref{table:1} and \ref%
{table:2}. Table \ref{table:1} contains the model results where few sources
of contributions can be identified\footnote{The N$\chi $QM results for the pion and pseudoscalar mesons exchanges
are taken from \cite{Dorokhov:2008pw}. The
scalar mesons contribution is corrected value of (from $0.39(0.04)$ \cite{Dorokhov:2012qa} to $0.34(0.48)$ \cite{DRZ14}).
The dynamical quark-loop contribution and the total result are taken from \cite{DRZ14}.}. In Table \ref{table:2} there are
the model results where only contribution of the light pseudoscalar mesons
is calculated.

To reduce the model dependence of various approaches, different constraints
on their parameter space are employed. One kind of important constraints on
the models follows from the phenomenology of the two-photon widths of the
pseudoscalar mesons $\Gamma \left( PS\rightarrow \gamma \gamma \right) $ and
their transition form factors $\mathrm{F}_{PS\gamma \gamma ^{\ast }}\left(
-M_{PS}^{2};0,q^{2}\right) $ first emphasized in \cite{Hayakawa:1995ps}.
Another set of constraints follows from the large momentum asymptotics for
the meson transition form factors \cite%
{Hayakawa:1995ps,Bijnens:1995cc,Knecht:2001qf} and for the total
light-by-light scattering amplitude considered in \cite%
{Melnikov:2003xd,Dorokhov:2008pw,Dorokhov:2012qa}, obtained using
perturbative QCD and reproduced within the N$\chi$QM.

\begin{figure}[h]
\resizebox{0.45\textwidth}{!}{\includegraphics{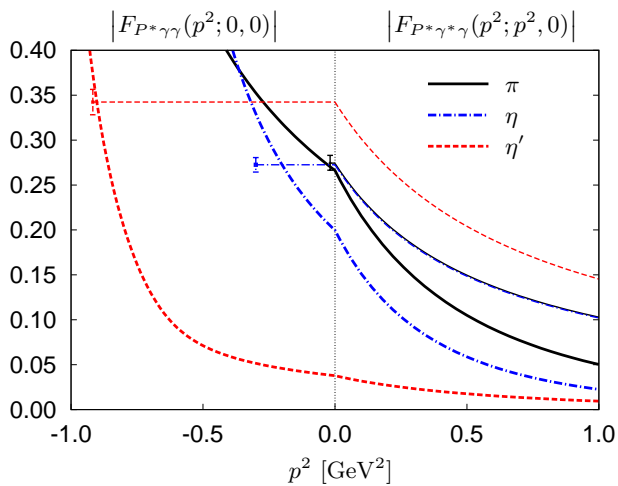}}
\caption{Plots of the $\protect\pi ^{0},\protect\eta $ and $\protect\eta %
^{\prime }$ vertices $F_{P^{\ast }\protect\gamma \protect\gamma }(p^{2};0,0)$
in the timelike region and $F_{P^{\ast }\protect\gamma ^{\ast }\protect%
\gamma }(p^{2};p^{2},0)$ in the spacelike region in N$\protect\chi $QM model
(thick lines) and VMD model (thin lines). The points with error bars
correspond to the physical points of the meson decays into two photons. The
VMD curves for $\protect\pi ^{0}$ and $\protect\eta $ are almost
indistinguishable.}
\label{fig:CompFF}
\end{figure}

In addition, the model amplitudes have to be consistent with the 4-momentum
conservation law. In practice, it means that the off-shell effects for
intermediate mesons should be taken into account \cite%
{Jegerlehner:2009ry,Dorokhov:2008pw,Dorokhov:2012qa}. For illustration of
this effect see Fig. \ref{fig:CompFF} from \cite{Dorokhov:2008pw}.

Finally, the model calculations should be tested by reproducing the
dispersion analysis result for the HVP contribution to $a_\mu$ as it
was done in \cite%
{deRafael:1993za,Pallante:1994ee,Pivovarov:2001mw,Dorokhov:2004ze,Greynat:2012ww}
and the known (semi)analytical results on the fermion-loop contributions to
the lepton AMM as it was done in \cite%
{Pivovarov:2001mw,Boughezal:2011vw,Dorokhov:2012qa,Greynat:2012ww}. In this
respect note, that in \cite{Kataev:2012kn} the CQM expression of \cite%
{Boughezal:2011vw} for the 4-loop HLbL contribution to the $a_{\mu }$ was
used for analytical evaluation of the first term of eighth-order $\left(
m_{\mu }/m_{\tau }\right) $ contribution to $a_{\mu }$ and $\left(
m_{e}/m_{\tau }\right) $ and $\left( m_{e}/m_{\mu }\right) $ contributions
to $a_{e}$. The analytical results turned out to be in good agreement with
the numerical results of calculations of these 8-th order massive
corrections, re-evaluated in the process of obtaining 10-th order
corrections to $a_{e}$ in \cite{Aoyama:2012wj} and complete 10-th order
corrections to $a_{\mu }$ \cite{Aoyama:2012wk}. In \cite{Kataev:2012kn} the
statement was made, that the comparison of the results obtained in \cite%
{Kataev:2012kn} with the numerical calculations of the 8-order massive
dependent corrections indicate that the numerical results are also sensitive
to higher order power-suppressed massive-dependent corrections. This
statement was confirmed by direct analytical calculations, performed
recently in \cite{Kurz:2013exa}. These QED calculations of \cite%
{Kataev:2012kn} and \cite{Kurz:2013exa} and the comparison with the results
of numerical 8-th order QED calculations of \cite{Aoyama:2012wj}
demonstrate, that the CQM calculations of 8-th order light-by-light
contributions to $a_{\mu }$ of \cite{Boughezal:2011vw} are correct.
\begin{table*}[th]
\centering%
\begin{tabular}{|l|l|l|l|l|l|l|l|}
\hline
Model & $\pi ^{0}$ & PS & S & AV & Quark & $\pi ,K-$ & Total \\
&  & $\left( \pi ^{0},\eta ,\eta ^{\prime }\right) $ & $\left( \sigma
,f_{0},a_{0}\right) $ &  & loop & loops &  \\ \hline
VMD (Hayakawa \cite{Hayakawa:1995ps}) & $5.74(0.36)$ & $8.27(0.64)$ &  & $%
0.17(0.10)$ & $0.97(1.11)$ & $-0.45(0.81)$ & $8.96(1.54)$ \\
ENJL (Bijnens \cite{Bijnens:1995cc}) & $5.58(0.05)$ & $8.5(1.3)$ & $%
-0.68(0.2)$ & $0.25(0.1)$ & $2.1(0.3)$ & $-1.9(1.3)$ & $8.3(3.2)$ \\
LMD+V (Knecht \cite{Knecht:2001qf}) & $5.8(1.0)$ & $8.3(1.2)$ &  &  &  &  & $%
8.0(4.0)$ \\
Q-box (Pivovarov \cite{Pivovarov:2001mw}) &  &  &  &  & $14.05$ &  & $14.05$
\\
LENJL (Bartos \cite{Bartos:2001pg}) & $8.18(1.65)$ & $9.55(1.7)$ & $%
1.23(0.24)$ &  &  &  & $10.77(1.68)$ \\
(LMD+V)$^{\prime }$(Melnikov \cite{Melnikov:2003xd}) & $7.65(1.0)$ & $%
11.4(1.0)$ &  & $2.2(0.5)$ &  & $0(10)$ & $13.6(0.25)$ \\
N$\chi $QM (Dorokhov \cite{Dorokhov:2008pw,Dorokhov:2012qa,DRZ14})
& $5.01(0.37)$
& $5.85(0.87)$
& $0.34(0.48)$ & 
& $11.0(0.9)$ &  
& $16.8(1.25)$ \\ 
oLMDV (Nyffeler \cite{Nyffeler:2009tw}) & $7.2(1.2)$ & $9.9(1.6)$ & $%
-0.7(0.2)$ & $2.2(0.5)$ & $2.1(0.3)$ & $-1.9(1.3)$ & $11.6(0.4)$ \\
DS (Goecke \cite{Fischer:2010iz}) & $5.75(0.69)$ & $8.07(1.2)$ &  &  & $%
10.7(0.2)$ &  & $18.8(0.4)$ \\
C$\chi $QM (Greynat \cite{Greynat:2012ww}) & $6.8(0.3)$ & $6.8(0.3)$ &  &  &
$8.2(0.6)$ &  & $15.0(0.3)$ \\ \hline
\end{tabular}
\vspace*{8pt}
\caption{Model estimates of the HLbL contribution to $a_\mu$
 from various sources obtained in different
works. All numbers are given in $10^{-10}$. The errors do not include the
systematic error of the models.
}
\label{table:1}
\end{table*}

\begin{table*}[th]
\centering%
\begin{tabular}{|l|l|l|}
\hline
Model & $\pi ^{0}$ & PS \\ \hline
Holography (Hong \cite{Hong:2009zw}) & $6.9$ & $10.7$ \\
Holography (Cappiello \cite{Cappiello:2010uy}) & $6.54(0.25)$ &  \\
R$\chi $T\ (Kampf \cite{Kampf:2011ty}) & $6.58(0.12)$ &  \\
R$\chi $T (Roig \cite{Roig:2014uja}) & $6.66(0.21)$ & $10.47(0.54)$ \\ \hline
\end{tabular}
\vspace*{8pt}
\caption{The HLbL contribution to $a_\mu$ from the mesonic
exchanges in the neutral pseudoscalar channel obtained in different works. 
All numbers are given in $10^{-10}$. }
\label{table:2}
\end{table*}


In the next part we discuss the HLbL contribution as it is calculated within the 
N$\chi $QM and show that,
within this framework, it might be possible to  realistically determine this
value to a sufficiently safe accuracy. We would like to discuss, how well this
model (see, e.g., \cite{Anikin:2000rq}) does in calculating $a_{\mu }^{%
\mathrm{HLbL}}$. Below, within the N$\chi $QM, we discuss in some details
the theoretical status of HLbL contributions to $a_\mu$ due to the exchange by light mesons and the
dynamical quark loop.

\section{HLbL contribution to the muon AMM within nonlocal chiral quark model%
}

\subsection{N$\protect\chi $QM dynamics}

The N$\chi $QM is an effective QCD inspired model that has a numerous applications for
description of low energy hadronic dynamics \cite{Anikin:2000rq}. We mention
only those applications that are related to the problem of hadronic
contributions to $a_\mu$. The two-point VV correlator has been
calculated in \cite{Dorokhov:2003kf} and used for calculations of
$a_\mu^{\mathrm{HVP}}$ \cite{Dorokhov:2004ze}. The three-point VAV
correlator has been studied in \cite{Dorokhov:2005pg} and used for
calculations of the hadronic photon-$Z$-boson vertex contribution to $a_\mu$ \cite{Dorokhov:2005ff}. 
The HLbL corrections due to light meson
exchanges and specific HVP corrections, where the virtual photon splits into
$\pi ^{0}\left( \sigma \right) $ and $\gamma $, was elaborated in \cite%
{Dorokhov:2008pw,Dorokhov:2012qa}. Note that the N$\chi $QM approach in many
ways similar to ENJL \cite{Bijnens:1995cc} and DS \cite{Fischer:2010iz}
models with, of course, subtle differences between all of them.

The Lagrangian of the $SU(3)\times SU(3)$ chiral quark model has the form
\begin{align}
\mathcal{L}& =\bar{q}(x)(i\hat{\partial}-m_{c})q(x)+\frac{G}{2}%
[J_{S}^{a}(x)J_{S}^{a}(x)+J_{PS}^{a}(x)J_{PS}^{a}(x)]  \notag \\
& -\frac{H}{4}%
T_{abc}[J_{S}^{a}(x)J_{S}^{b}(x)J_{S}^{c}(x)-3J_{S}^{a}(x)J_{PS}^{b}(x)J_{PS}^{c}(x)],
\label{33model}
\end{align}%
where $q\left( x\right) $ are the quark fields, $m_{c}$ $\left(
m_{u}=m_{d}\neq m_{s}\right) $ is the diagonal matrix of the quark current
masses, $G$ and $H$ are the four- and six-quark coupling constants. Second
line in the Lagrangian represents the Kobayashi--Maskawa--t`Hooft
determinant vertex with the structural constant
\begin{equation*}
T_{abc}=\frac{1}{6}\epsilon _{ijk}\epsilon _{mnl}(\lambda _{a})_{im}(\lambda
_{b})_{jn}(\lambda _{c})_{kl},
\end{equation*}%
where $\lambda _{a}$ are the Gell-Mann matrices for $a=1,..,8$ and $\lambda
_{0}=\sqrt{2/3}I$.

The nonlocal structure of the model is introduced via the nonlocal quark
currents
\begin{equation}
J_{M}^{a}(x)=\int d^{4}x_{1}d^{4}x_{2}\,f(x_{1})f(x_{2})\,\bar{q}%
(x-x_{1})\,\Gamma _{M}^{a}q(x+x_{2}),  \label{JaM}
\end{equation}%
where $M=S$ for the scalar and $M=PS$ for the pseudoscalar channels, $\Gamma
_{{S}}^{a}=\lambda ^{a}$, $\Gamma _{{PS}}^{a}=i\gamma ^{5}\lambda ^{a}$ and $%
f(x)$ is a form factor with the nonlocality parameter $\Lambda $ reflecting
the nonlocal properties of the QCD vacuum.

The model (\ref{33model}) can be bosonized using the stationary phase
approximation which leads to the system of gap equations for the dynamical
quark masses $m_{d,i}$%
\begin{equation}
m_{d,i}+GS_{i}+\frac{H}{2}S_{j}S_{k}=0,  \label{GapEqs}
\end{equation}%
with $i=u,d,s$ and $j,k\neq i,$ and $S_{i}$ is the quark loop integral
\begin{equation*}
S_{i}=-8N_{c}\int \frac{d_{E}^{4}k}{(2\pi )^{4}}\frac{%
f^{2}(k^{2})m_{i}(k^{2})}{D_{i}(k^{2})},
\end{equation*}%
where $m_{i}(k^{2})=m_{c,i}+m_{d,i}f^{2}(k^{2})$, $%
D_{i}(k^{2})=k^{2}+m_{i}^{2}(k^{2})$ is the dynamical quark propagator
obtained by solving the DS equation, $f(k^{2})$ is the nonlocal form factor
in the momentum representation.
\begin{figure}[h]
 \includegraphics[height=1.2cm]{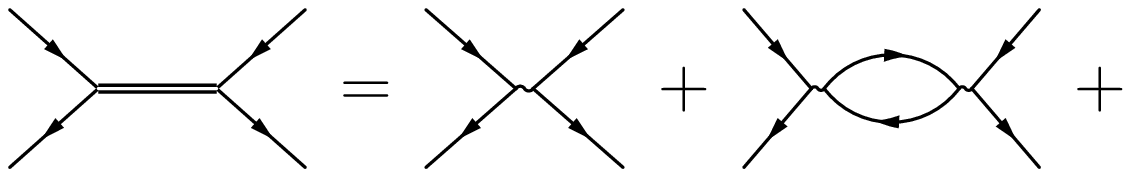}
\includegraphics[height=1.2cm]{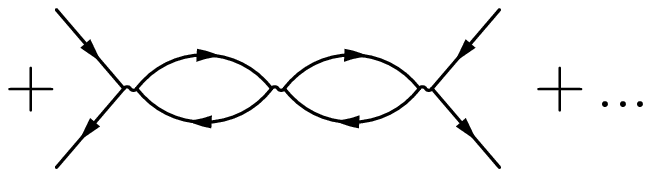}
 \vspace*{8pt}
\caption{The Bethe-Salpeter equation for meson propagators. The 4-quark
crosses are for the interaction term (\protect\ref{33model}).}
\label{Fig: BS}
\end{figure}

The quark-meson vertex functions and the meson masses can be found from the
Bethe-Salpeter equation Fig. \ref{Fig: BS}. For the separable interaction (%
\ref{33model}) the quark-antiquark scattering matrix in each ($PS$ or $S$)
channels becomes
\begin{align}
& \mathbf{T}=\hat{\mathbf{T}}(p^{2})\delta ^{4}\left(
p_{1}+p_{2}-(p_{3}+p_{4})\right) \prod\limits_{i=1}^{4}f(p_{i}^{2}),  \notag
\\
& \hat{\mathbf{T}}(p^{2})=i\gamma _{5}\lambda _{k}\left( \frac{1}{-\mathbf{G}%
^{-1}+\mathbf{\Pi }(p^{2})}\right) _{kl}i\gamma _{5}\lambda _{l},
\end{align}%
where $p_{i}$ are the momenta of external quark lines, $\mathbf{G}$ and $%
\mathbf{\Pi }(p^{2})$ are the corresponding matrices of the four-quark
coupling constants and the polarization operators of mesons ($%
p=p_{1}+p_{2}=p_{3}+p_{4}$). The meson masses can be found from the zeros of
determinant $\mathrm{det}(\mathbf{G}^{-1}-\mathbf{\Pi }(-M^{2}))=0.$ The $%
\hat{\mathbf{T}}$-matrix for the system of mesons in each neutral channel
can be expressed as
\begin{equation}
\hat{\mathbf{T}}_{ch}(P^{2})=\sum_{M_{ch}}\frac{\overline{V}%
_{M_{ch}}(P^{2})\otimes V_{M_{ch}}(P^{2})}{-(P^{2}+\mathrm{M}_{{M_{ch}}}^{2})},  \label{Tch}
\end{equation}%
where $\mathrm{M}_{M}$ are the meson masses, $V_{M}(P^{2})$ are the vertex
functions $\left( \overline{V}_{M}(p^{2})=\gamma ^{0}V_{M}^{\dag
}(P^{2})\gamma ^{0}\right) $. The sum in (\ref{Tch}) is over full set of
light mesons: $(M_{PS}={\pi ^{0},\eta ,\eta ^{\prime }})$ in the
pseudoscalar channel and $(M_{S}={a_{0}(980),f_{0}(980),\sigma })$ in the
scalar one.

\subsection{External photon fields}

The gauge-invariant interaction with an external photon field $V_{\mu }^{a}$
can be introduced through the Schwinger phase factor%
\begin{equation}
q\left( y\right) \rightarrow Q\left( x,y\right) =\mathcal{P}\exp \left\{
i\int_{x}^{y}dz^{\mu }V_{\mu }^{a}\left( z\right) T^{a}\right\} q\left(
y\right) .  \label{SchwPhF}
\end{equation}%
Then, apart from the kinetic term, the additional terms in the nonlocal
interaction are generated via%
\begin{eqnarray}
J_{M}^{a}(x)\rightarrow J_{M}^{a}(x)=\int
d^{4}x_{1}d^{4}x_{2} \,f(x_{1})f(x_{2}) \times\nonumber\\
\,\times\,\overline{Q}(x-x_{1},x)\,\Gamma
_{M}^{a}Q(x,x+x_{2}),  \label{JaMgauge}
\end{eqnarray}%
which induces the quark-antiquark--$n$-photon vertices. Additionally, there
appear the meson--quark-anti-quark--$n$-photon vertices. The following
equations are used for obtaining the nonlocal vertices \cite%
{Mandelstam:1962mi}%
\begin{eqnarray}
\frac{\partial }{\partial y^{\mu }}\int_{x}^{y}dz^{\nu }V_{\nu }\left(
z\right) =V_{\mu }\left( y\right) , \nonumber \\
\delta ^{\left( 4\right) }\left(
x-y\right) \int_{x}^{y}dz^{\nu }V_{\nu }\left( z\right) =0.
\label{GaugeRules}
\end{eqnarray}

\begin{figure}[h]
\begin{center}
\centerline{\begin{tabular*}{\columnwidth}{@{\extracolsep{\fill}}cccc}
\resizebox{0.15\textwidth}{!}{\includegraphics{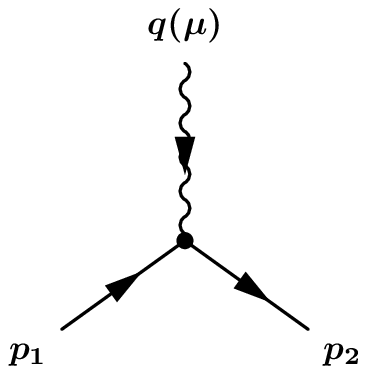}} & \resizebox{0.15\textwidth}{!}{\includegraphics{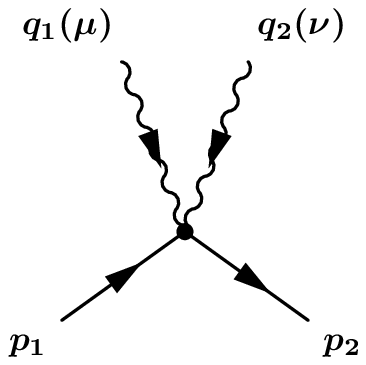}} &  &  \\
(a) & (b) &  &
\end{tabular*}}
\end{center}
\caption{The quark-photon vertex $\mathrm{\Gamma }_{\protect\mu }^{\left(
1\right) }\left( q\right) $, Eq. (\protect\ref{GamTot}) (a), and the
quark-2-photon vertex $\mathrm{\Gamma }_{\protect\mu \protect\nu %
}^{(2)}\left( q_{1},q_{2}\right) $, Eq. (\protect\ref{DGam2}) (b). }
\label{fig:VerticesPhot-n}
\end{figure}

As an example, the quark-antiquark vertices with one-photon (Fig. \ref%
{fig:VerticesPhot-n}a) and two-photon (Fig. \ref{fig:VerticesPhot-n}b)
insertions are%
\begin{equation}
\Gamma _{\mu }^{\left( 1\right) }=\gamma _{\mu }+\Delta \Gamma _{\mu
}^{\left( 1\right) }\left( q_{1}\right) ,  \label{GamTot}
\end{equation}%
\begin{eqnarray}
&&\Delta \Gamma _{\mu }^{\left( 1\right) }\left( q_{1}\right) =-\left(
p_1+k_{1}\right) _{\mu }m^{\left( 1\right) }\left( p_1,k_{1}\right) ,
\label{DGam1} \\
&&\Gamma _{\mu \nu }^{\left( 2\right) }\left( q_{1},q_{2}\right) =2g_{\mu
\nu }m^{\left( 1\right) }\left( p_1,k_{12}\right)+   \nonumber\\
&&\quad\left( p_1+k_{1}\right) _{\mu }\left( k_{1}+k_{12}\right) _{\nu }m^{\left(
2\right) }\left( p_1,k_{1},k_{12}\right)+ \label{DGam2}\\
&&\quad\left( p_1+k_{2}\right) _{\nu }\left(
k_{2}+k_{12}\right) _{\mu }m^{\left( 2\right) }\left( p_1,k_{2},k_{12}\right)
,
\notag
\end{eqnarray}%
where the finite-difference derivatives are introduced%
\begin{eqnarray}
f^{\left( 1\right) }\left( a,b\right) &=&\frac{f\left( a+b\right) -f\left(
b\right) }{\left( a+b\right) ^{2}-b^{2}},  \label{FDD1} \\
f^{\left( 2\right) }\left( a,b,c\right) &=&\frac{f^{\left( 1\right) }\left(
a,b\right) -f^{\left( 2\right) }\left( a,c\right) }{\left( a+b\right)
^{2}-\left( a+c\right) ^{2}},
  \label{FDD2}
\end{eqnarray}%
In (\ref{GamTot}-\ref{DGam2}), $p_1$ is the momentum of incoming quark, $q_{i}$
are the momenta of incoming photons and $k_{1}=k+q_{1},$ $%
k_{ij...k}=p_1+q_{i}+q_{j}+...+q_{k}$. The vertex $\Gamma _{\mu }$ satisfies
the Ward-Takahashi identity for dynamical quarks%
\begin{equation}
q_{1}\Gamma _{\mu }^{\left( 1\right) }=S^{-1}\left( p_1+q_{1}\right)
-S^{-1}\left( p_1\right) ,  \label{WTloc}
\end{equation}%
with
\begin{equation}
S^{-1}\left( p\right)=\hat p -m(p) ,  \label{S-1}
\end{equation}
and for the multi-photon nonlocal vertices one has%
\begin{eqnarray}
&&q_{1}^{\mu }\Delta \Gamma _{\mu }^{\left( 1\right) }\left( q_{1}\right)
=m\left( p_1\right) -m\left( k_{1}\right) ,  \label{WT1} \\
&&q_{1}^{\mu }q_{2}^{\nu }\Gamma _{\mu \nu }^{\left( 2\right) }\left(
q_{1},q_{2}\right) =m\left( p_1\right) +m\left( k_{12}\right) -m\left(
k_{1}\right) -m\left( k_{2}\right) .\nonumber
\end{eqnarray}%

\subsection{Hadronic Light-by-light contribution to $a_\mu$ within N$%
\protect\chi $QM}
\begin{figure*}[tbh]
\centerline{\includegraphics[width=12.7cm]{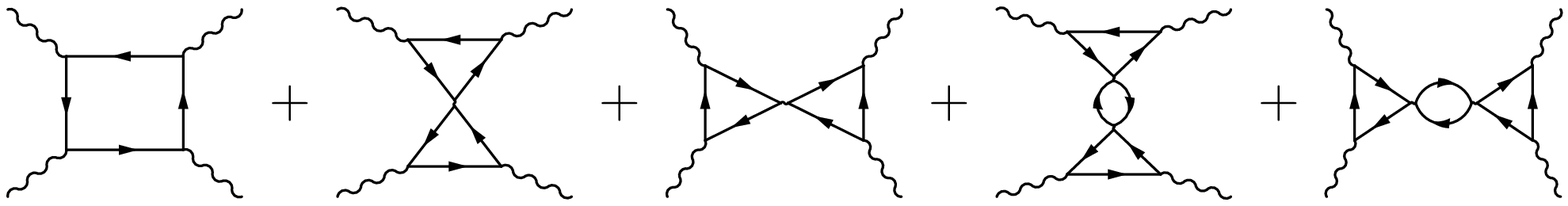}} \centerline{%
\includegraphics[width=7.62cm]{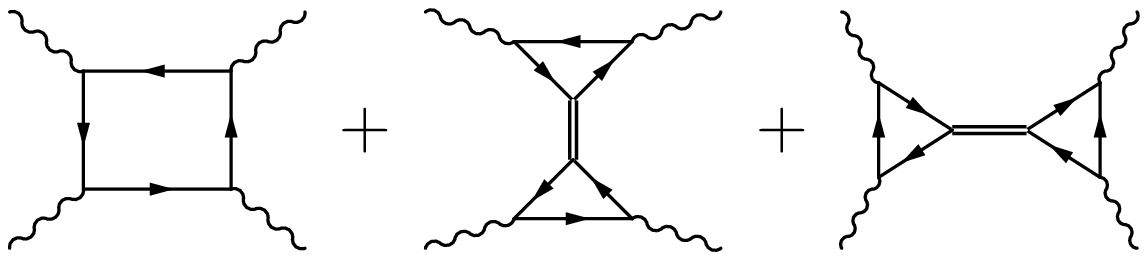}} \vspace*{8pt}
\caption{A schematic illustration for the diagrams contributing to the
four-rank polarization tensor to the leading in $1/N_{c}$ order. The
nonlocal multi-photon vertices are not shown for simplicity, see Fig.
\protect\ref{Fig: BoxCont}.}
\label{BoxAll}
\end{figure*}
The basic element for the calculations of $a_\mu^{\mathrm{HLbL}}$ is the fourth-rank light quark hadronic vacuum polarization tensor%
\begin{align}
& \mathrm{\Pi }_{\mu \nu \lambda \rho }(q_{1},q_{2},q_{3})=\int
d^{4}x_{1}\int d^{4}x_{2}\int
d^{4}x_{3}\times   \label{P4gamma} \\
& \quad \times e^{i(q_{1}x_{1}+q_{2}x_{2}+q_{3}x_{3})}\left\langle 0|T(j_{\mu }(x_{1})j_{\nu
}(x_{2})j_{\lambda }(x_{3})j_{\rho }(0))|0\right\rangle ,\nonumber
\end{align}%
where $j_{\mu }(x)$ are the light quark electromagnetic currents and $%
\left\vert 0\right\rangle $ is the QCD vacuum state. The muon AMM can be
extracted by using the projection \cite{Brodsky:1967sr}
\begin{equation*}
a_{\mu }^{\mathrm{HLbL}}=\frac{1}{48m_{\mu }}\mathrm{Tr}\left( (\hat{p}%
+m_{\mu })[\gamma ^{\rho },\gamma ^{\sigma }](\hat{p}+m_{\mu })\mathrm{\Pi }%
_{\rho \sigma }(p,p)\right) ,
\end{equation*}%
where
\begin{align}
& \mathrm{\Pi }_{\rho \sigma }(p^{\prime },p)=-ie^{6}\int \frac{d^{4}q_{1}}{%
(2\pi )^{4}}\int \frac{d^{4}q_{2}}{(2\pi )^{4}}\frac{1}{%
q_{1}^{2}q_{2}^{2}(q_{1}+q_{2}-k)^{2}}\times  \nonumber\\
& \quad  \times \gamma ^{\mu }\frac{\hat{p}^{\prime }-\hat{q}%
_{1}+m_{\mu }}{(p^{\prime }-q_{1})^{2}-m_{\mu }^{2}}\gamma ^{\nu }\frac{\hat{%
p}-\hat{q}_{1}-\hat{q}_{2}+m_{\mu }}{(p-q_{1}-q_{2})^{2}-m_{\mu }^{2}}\gamma
^{\lambda }\times   \nonumber\\
& \quad  \times \frac{\partial }{\partial k^{\rho }}\mathrm{\Pi }_{\mu \nu
\lambda \sigma }(q_{1},q_{2},k-q_{1}-q_{2}), \label{P4gamProject}
\end{align}%
with $m_{\mu }$ is the muon mass, $k_{\mu }=(p^{\prime }-p)_{\mu }$ and it
is necessary to make the limit $k_{\mu }\rightarrow 0$.

In the N$\chi $QM, the tensor $\mathrm{\Pi }_{\mu \nu \lambda \rho }$ is
represented in the leading in $1/N_{c}$ order by the chain of diagrams
schematically depicted in Fig. \ref{BoxAll}. In the higher order
contributions, the $1/N_{c}$ suppression factor coming from the four-quark
interaction (\ref{33model}) is compensated by the $N_{c}$ factor from the
color trace of the quark loop. This infinite series of quark loop diagrams
is summed up leading to the quark box and the diagrams with light meson
exchanges. The double chain summation generates the meson loop contributions
which are, however, suppressed by $1/N_{c}$ factor.

\begin{figure}[tbh]
\begin{center}
\begin{tabular*}{\columnwidth}{@{}ccc}
\raisebox{-0.5\height}{\resizebox{0.13\textwidth}{!}{%
\includegraphics{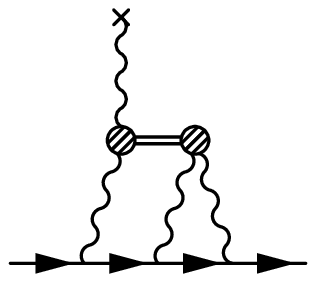}}} & \raisebox{-0.5\height}{\resizebox{0.13%
\textwidth}{!}{\includegraphics{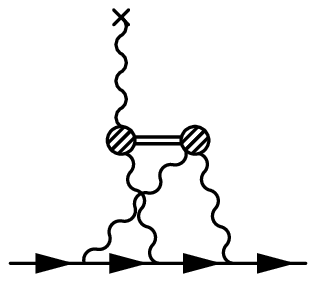}}} & \raisebox{-0.5\height}{%
\resizebox{0.13\textwidth}{!}{\includegraphics{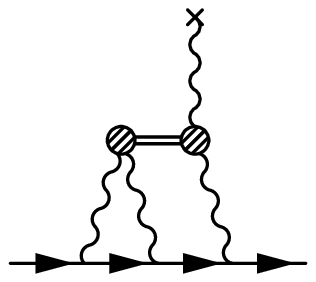}}} \\
(a) & (b) & (c)%
\end{tabular*}%
\end{center}
\caption{The HLbL contribution from the intermediate light meson exchanges.}
\label{fig:LbLmes}
\end{figure}

The HLbL contribution due to exchange of pseudoscalar (PS) and scalar (S)
mesons (Fig. \ref{fig:LbLmes}) was elaborated in \cite{Dorokhov:2008pw}. The
vertices containing the virtual (off-shell) meson $M$ with momentum $p$ and
two virtual photons with momenta $q_{1,2}$ and the polarization vectors $%
\epsilon _{1,2}$ can be written as \cite{Bartos:2001pg}
\begin{equation}
\mathcal{A}\left( \gamma _{\left( q_{1},\epsilon _{1}\right) }^{\ast }\gamma
_{\left( q_{2},\epsilon _{2}\right) }^{\ast }\rightarrow M_{\left( p\right)
}^{\ast }\right) =e^{2}\epsilon _{1}^{\mu }\epsilon _{2}^{\nu }\Delta
_{M^{\ast }\gamma ^{\ast }\gamma ^{\ast }}^{\mu \nu }\left(
p;q_{1},q_{2}\right) ,  \label{MGGvert}
\end{equation}%
where for the pseudoscalar mesons
\begin{equation}
\Delta _{PS^{\ast }\gamma ^{\ast }\gamma ^{\ast }}^{\mu \nu }\left(
p;q_{1},q_{2}\right) =-i\varepsilon _{\mu \nu \rho \sigma }q_{1}^{\rho
}q_{2}^{\sigma }\mathrm{F}_{PS^{\ast }\gamma ^{\ast }\gamma ^{\ast }}\left(
p^{2};q_{1}^{2},q_{2}^{2}\right) ,  \label{PiGGvert}
\end{equation}%
and for the scalar mesons
\begin{eqnarray}
&&\Delta _{S^{\ast }\gamma ^{\ast }\gamma ^{\ast }}^{\mu \nu }\left(
p;q_{1},q_{2}\right) =
\mathrm{A}_{S^{\ast }\gamma ^{\ast }\gamma ^{\ast }}\left( p^{2};q_{1}^{2},q_{2}^{2}\right) T_{A}^{\mu \nu }(q_{1},q_{2})\nonumber \\%
&&\quad\quad+\mathrm{B}_{S^{\ast }\gamma ^{\ast }\gamma ^{\ast }}(p^{2};q_{1}^{2},q_{2}^{2})T_{B}^{\mu \nu }(q_{1},q_{2}),
\label{SigmaGGvert}
\end{eqnarray}%
and the Lorentz structures are
\begin{eqnarray}
&&\quad T_{A}^{\mu \nu }(q_{1},q_{2})=\left( g^{\mu \nu
}(q_{1}q_{2})-q_{1}^{\nu }q_{2}^{\mu }\right) ,  \label{TA} \\
&&\quad T_{B^{\prime }}^{\mu \nu }(q_{1},q_{2})=
{
\left(
q_{1}^{2}q_{2}^{\mu }-(q_{1}q_{2})q_{1}^{\mu }\right) \left(
q_{2}^{2}q_{1}^{\nu }-(q_{1}q_{2})q_{2}^{\nu }\right) }
\nonumber
\end{eqnarray}%
and $p=q_{1}+q_{2}$. The subject of model calculations \cite{Dorokhov:2008pw}
is to get the (PS/S)$^{\ast }$V$^{\ast }$V$^{\ast }$ vertex functions $%
\mathrm{F}_{PS},\mathrm{A}_{S},\mathrm{B}_{S}$.
\begin{figure*}[t]
\begin{center}
\begin{tabular*}{\textwidth}{ccccccccccccc}
\raisebox{-0.5\height}{\resizebox{!}{0.08\textheight}{%
\includegraphics{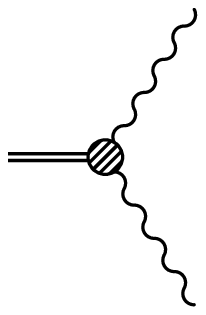}}}&\raisebox{-0.5\height}{=}&\raisebox{-0.5%
\height}{\resizebox{!}{0.08\textheight}{\includegraphics{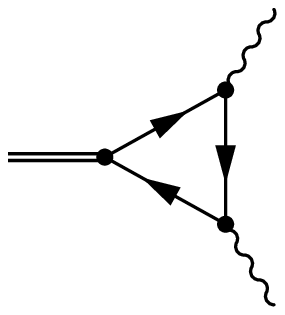}}} & %
\raisebox{-0.5\height}{+} & \raisebox{-0.5\height}{\resizebox{!}{0.08%
\textheight}{\includegraphics{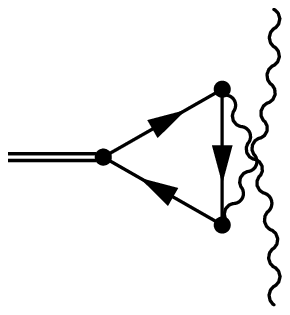}}} & \raisebox{-0.5\height}{+} & %
\raisebox{-0.5\height}{\resizebox{!}{0.08\textheight}{%
\includegraphics{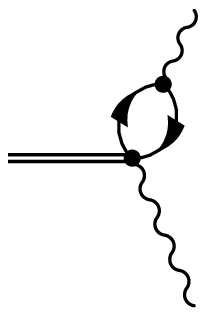}}} & \raisebox{-0.5\height}{+} & \raisebox{-0.5%
\height}{\resizebox{!}{0.08\textheight}{\includegraphics{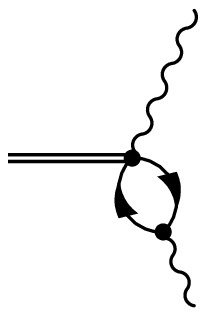}}} & %
\raisebox{-0.5\height}{+} & \raisebox{-0.5\height}{\resizebox{!}{0.08%
\textheight}{\includegraphics{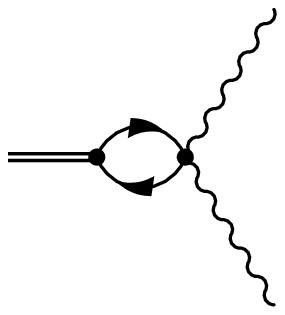}}} & \raisebox{-0.5\height}{+} & %
\raisebox{-0.5\height}{\resizebox{!}{0.08\textheight}{%
\includegraphics{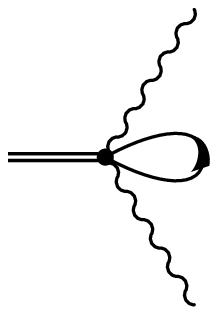}}} \\
(a) &  & (b) &  & (c) &  & (d) &  & (e) &  & (f) &  & (g)%
\end{tabular*}%
\end{center}
\caption{The diagrams for the photon-meson transition form factors $\mathrm{F%
}_{PS^{\ast }\protect\gamma ^{\ast }\protect\gamma ^{\ast }}\mathrm{%
,A_{S^{\ast }\protect\gamma ^{\ast }\protect\gamma ^{\ast }},B_{S^{\ast }%
\protect\gamma ^{\ast }\protect\gamma ^{\ast }}}$. In the case of
pseudoscalar mesons, the diagrams (d-g) give zero contributions due to
chirality considerations.}
\label{fig:SggSimp}
\end{figure*}

The expression for $a_\mu^{\mathrm{HLbL}}$ from the light
meson exchanges can be written in a three-dimensional integral
representation as follows
\begin{eqnarray}
&&a_{\mu }^{\mathrm{HLbL},\mathrm{MesExch}}=-\frac{2\alpha ^{3}}{3\pi ^{2}}%
\int\limits_{0}^{\infty }dQ_{1}^{2}\int\limits_{0}^{\infty
}dQ_{2}^{2}\int\limits_{-1}^{1}dt\sqrt{1-t^{2}}\times \nonumber\\
&&\times\frac{1}{Q_{3}^{2}}\sum_{M}%
\left[ 2\frac{\mathcal{N}_{1}^{M}(Q_{1}^{2},Q_{2}^{2},Q_{3}^{2})}{Q_{2}^{2}+%
\mathrm{M}_{M}^{2}}+\frac{\mathcal{N}_{2}^{M}(Q_{1}^{2},Q_{3}^{2},Q_{2}^{2})%
}{(Q_{3}^{2}+\mathrm{M}_{M}^{2})}\right] , \nonumber \\
&&\quad \mathcal{N}_{\mathbf{1,2}}^{PS}(Q_{1}^{2},Q_{2}^{2},Q_{3}^{2})=%
\mathrm{F}_{PS^{\ast }\gamma ^{\ast }\gamma ^{\ast }}\left(
Q_{2}^{2};Q_{2}^{2},0\right)\times \nonumber \\
&&\quad\quad\times \mathrm{F}_{PS^{\ast }\gamma ^{\ast }\gamma
^{\ast }}\left( Q_{2}^{2};Q_{1}^{2},Q_{3}^{2}\right) \mathrm{Tps}_{\mathbf{%
1,2}}, \label{amuQ1Q2}  \\
&&\quad \mathcal{N}_{1,2}^{S}(Q_{1}^{2},Q_{2}^{2},Q_{3}^{2})=
\mathrm{A%
}_{S^{\ast }\gamma ^{\ast }\gamma ^{\ast }}\left(
Q_{2}^{2};Q_{2}^{2},0\right) \times
\notag \nonumber  \\
&&\quad\quad \times\biggl(
\mathrm{A}_{S^{\ast }\gamma ^{\ast }\gamma ^{\ast }}\left(
Q_{2}^{2};Q_{1}^{2},Q_{3}^{2}\right) \mathrm{Ts}_{\mathbf{1,2}}^{\mathrm{AA}%
} \nonumber \\
&&\quad\quad\quad\quad
+\frac{1}{2}\mathrm{B}_{S^{\ast }\gamma ^{\ast }\gamma ^{\ast }}\left(
Q_{2}^{2};Q_{1}^{2},Q_{3}^{2}\right) \mathrm{Ts}_{\mathbf{1,2}}^{\mathrm{AB}%
}\biggr) ,  \notag
\end{eqnarray}%
where $Q_{3}=-\left( Q_{1}+Q_{2}\right) $, $t=\left( Q_{1}Q_{2}\right)
/\left( \left\vert Q_{1}\right\vert \left\vert Q_{2}\right\vert \right) $.
The universal kinematic factors $%
\mathrm{Tps}_{\mathbf{1,2}}$ and $\mathrm{Ts}_{\mathbf{1,2}}$ obtained after
averaging over the directions of muon momentum $p$ can be found in \cite%
{Jegerlehner:2009ry} and \cite{Dorokhov:2008pw}, correspondingly. We would
like to stress that the integral representation $\left( \ref{amuQ1Q2}\right)
$ is valid for any form factors $\mathrm{F,A,B}$.

\begin{figure}[t]
\resizebox{0.45\textwidth}{!}{  \includegraphics{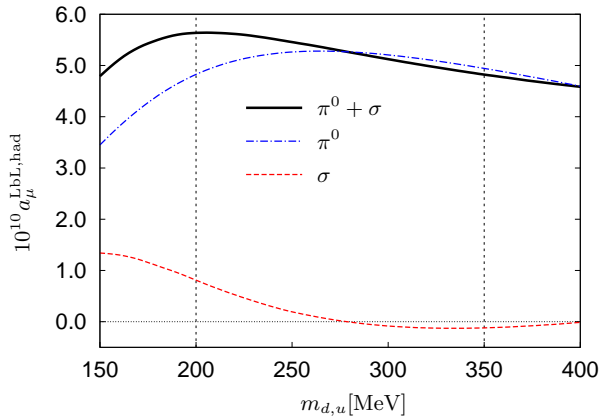}
}
\caption{ The HLbL contribution to $a_\mu$ from the neutral pion and $%
\protect\sigma $ exchanges as a function of the dynamical quark mass.
The lower line correspond to the $\sigma $ contribution, the $\pi ^{0}$ contribution is in the middle, and the upper line is the combined contribution.
Vertical thin dashed lines denote the interval of dynamical quark
masses used for the estimation of the error band for $a_{\protect\mu }^{%
\mathrm{LbL}}$. }
\label{fig:PionLbL}
\end{figure}

For numerical estimations in the N$\chi $QM we use the Gaussian nonlocal
form factor%
\begin{equation}
f\left( k^{2}\right) =\mathrm{exp}\left( -k^{2}/2\Lambda ^{2}\right) .  \label{fk}
\end{equation}%
Concerning the model parameters, the dynamical quark mass $m_{d}$ is taken in the
typical interval 200--350 MeV and then other parameters (the current quark
masses $m_{c}$ and the nonlocality parameter $\Lambda $) are fitted by the
pion mass and the two-photon decay constant in correspondence with the pion
lifetime given within the error range of PDG in \cite{Beringer:1900zz}. The
results are given in Table \ref{table:1}. Within the N$\chi $QM, we found
that the pseudoscalar meson contributions to $a_\mu$ are systematically
lower then the results obtained in the other works. The full kinematic
dependence of the vertices on the pion virtuality\footnote{%
Later, this dependence has also been studied in \cite{Fischer:2010iz}.}
diminishes the result by about 20-30\% as compared to the case where this
dependence is neglected. For $\eta $ and $\eta ^{\prime }$ mesons the
results are reduced by factor about 3 in comparison with the results
obtained in other models where the kinematic dependence was neglected (see
Fig. \ref{fig:CompFF}). The scalar mesons contribution is small and
partially compensates model dependence of the pseudoscalar contribution
(Fig. \ref{fig:PionLbL}).

The N$\chi $QM estimate for the contribution of the dynamical quark box to
$a_\mu$\footnote{%
The details of the calculations of the box diagram in N$\chi $QM will be presented
elsewhere \cite{DRZ14}.}, including the contact terms (see Fig. \ref{Fig:
BoxCont}), is given in Table \ref{table:1}. One can see that the momentum
dependent dynamical mass leads to increasing of contribution of diagram with
pure local quark--anti-quark--photon vertices in comparison with constant
quark mass. This behavior can be expected since $m(k^{2}\rightarrow \infty
)\rightarrow m_{c}$. The specific feature of these calculations is that
there is strong compensation between the contributions from the box diagram
with dynamical quarks and local vertices $\gamma ^{\mu },$ the box diagrams
with at least one nonlocal vertex $\Delta \Gamma _{\mu }^{\left( 1\right)
}(q)$ and all other types of nonlocal diagrams with contact vertices.

\begin{figure*}[t]
\centerline{\includegraphics[width=12.7cm]{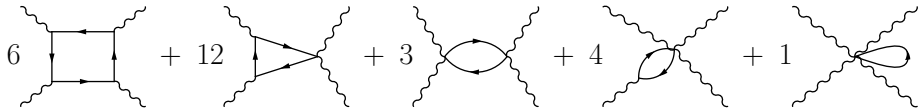}} \vspace*{8pt}
\caption{Contact terms which are gave contribution to $\mathrm{\Pi }_{%
\protect\mu \protect\nu \protect\lambda \protect\rho }(q_{1},q_{2},q_{3})$.
Numbers in front of diagrams are the degeneracy factors. }
\label{Fig: BoxCont}
\end{figure*}

\section{Conclusions}

We briefly discussed the current status of the experimental and theoretical
results on the electron and muon AMM, $g-2$. These quantities, measured and
calculated with very high accuracy, provide a very hard test of the SM. In
particular, the electron AMM tests QED at very short distances now
provides the best determination of the fine structure constant $\alpha $.
The muon AMM is much more sensitive to the effects of the Physics beyond the SM.
Presently, there is a mismatch between the latest experimental BNL
measurements and SM calculations at the level of 3-4 $\sigma $. It is the
largest deviation in elementary particle physics from the SM predictions. This
may be an evidence for the existence of new interactions and stringently
constrains the parametric space of hypothetical interactions extending the
SM. Nowdays, the interest in this problem became lively again in view of the preparation of new
Fermilab and J-PARC experiments planning to achieve a measurement precision at the
0.14 ppm level. On the other hand, the biggest theoretical uncertainty is
due to the hadronic part of $a_{\mu }^{\mathrm{SM}}$, especially from HVP
and HLbL. In this work we considered the latest achievements in
phenomenological and model approaches to estimate the leading and
next-to-leading order hadronic corrections to $a_{\mu }^{\mathrm{SM}}$.
Further studies are needed in order to get a better control over the hadronic corrections and reach
a precision of calculations comparable to or better than the
experimental one.

We thank Yu.M. Bystritskiy, A.L. Kataev, N.I. Kochelev, E.A. Kuraev,
V.P. Lomov, A.~Nyffeler, H.-P. Pavel, A.A. Pivovarov for critical remarks and illuminating discussions. This work is
supported in part by the Russian Foundation for Basic Research (projects No.
11-02-00112 and 12-02-31874).

\qquad\qquad

\end{document}